\documentclass[aps, pra, twocolumn, floatfix, showpacs, reprint, superscriptaddress]{revtex4-1}

\usepackage{graphicx}
\usepackage{amsmath, amsfonts, amssymb, bm}

\usepackage{amstext}
\usepackage{mathrsfs}
\usepackage{color}

\newcommand{\corr}[1]{{\color{black}#1}}

\begin{document}
\title{Improved local-constant-field approximation for strong-field QED codes}
\author{A.~Di~Piazza}
\email{dipiazza@mpi-hd.mpg.de}
\affiliation{Max-Planck-Institut f\"ur Kernphysik, Saupfercheckweg 1, D-69117 Heidelberg, Germany}
\author{M.~Tamburini}
\email{mtambu@mpi-hd.mpg.de}
\affiliation{Max-Planck-Institut f\"ur Kernphysik, Saupfercheckweg 1, D-69117 Heidelberg, Germany}
\author{S.~Meuren}
\affiliation{Department of Astrophysical Sciences, Princeton University, Princeton, New Jersey 08544, USA}
\author{C.~H.~Keitel}
\affiliation{Max-Planck-Institut f\"ur Kernphysik, Saupfercheckweg 1, D-69117 Heidelberg, Germany}

\date{\today}

\begin{abstract}
The local-constant-field approximation (LCFA) is an essential theoretical tool for investigating strong-field QED phenomena in background electromagnetic fields with complex spacetime structure. In our previous work [Phys.~Rev.~A~\textbf{98}, 012134 (2018)] we have analyzed the shortcomings of the LCFA in nonlinear Compton scattering at low emitted photon energies for the case of a background plane-wave field. Here, we generalize that analysis to background fields, which can feature a virtually arbitrary spacetime structure. In addition, we provide an explicit and simple implementation of an improved expression of the nonlinear Compton scattering differential probability that solves the main shortcomings of the standard LCFA in the infrared region, and is suitable for background electromagnetic fields with arbitrary spacetime structure such as those occurring in particle-in-cell simulations. Finally, we carry out a systematic procedure to calculate the probability of nonlinear Compton scattering per unit of emitted photon light-cone energy and of nonlinear Breit-Wheeler pair production per unit of produced positron light-cone energy beyond the LCFA in a plane-wave background field, which allows us to identify the limits of validity of this approximation quantitatively.
\end{abstract}

\maketitle

\section{Introduction}

High-energy processes in conventional accelerators typically involve only a few particles. However, when elementary particles interact in the presence of an intense background electromagnetic field, the relatively high density of photons in the field and their coherent behavior make elementary processes occur with the participation of a large number of such photons. Multiphoton effects in strong-field QED (SFQED) are controlled by Lorentz- and gauge-invariant parameters, which depend on the structure of the external electromagnetic field \cite{Landau_b_4_1982}. As a prominent example, intense laser radiation can be employed to investigate QED processes in the nonlinear regime. By considering processes involving the lightest charged particles, electrons and positrons, the parameter controlling nonlinear effects in the laser field amplitude is the so-called classical intensity parameter $\xi_0=|e|E_0/m\omega_0$. Here, $e<0$ and $m$ are the electron charge and mass, respectively, and $E_0$ and $\omega_0$ are the laser field amplitude and central angular frequency, respectively (units with $\hbar = c = 4\pi\epsilon_0 = 1$ are employed throughout the paper) \cite{Mitter_1975, Ritus_1985, Ehlotzky_2009, Reiss_2009, Di_Piazza_2012, Dunne_2014}. Present infrared and optical lasers with $\omega_0 \sim 1\;\text{eV}$ routinely exceed the threshold $\xi_0=1$, which corresponds to an intensity of the order of $10^{18}\;\text{W/cm$^2$}$ \cite{Yanovsky_2008} and future facilities aim at values of $\xi_0$ beyond one hundred, where nonlinear effects start becoming important also in the motion of protons \cite{APOLLON_10P, CoReLS, ELI, XCELS}. In fact, the physical origin of the nonlinear effects controlled by the parameter $\xi_0$ can ultimately be ascribed to the fact that in laser beams characterized by $\xi_0\gtrsim 1$ the magnetic force of the laser field on an electron, which depends on the electron's velocity, becomes comparable to the electric force.

There is another class of nonlinear effects that are intrinsically quantum mechanical and are controlled by another parameter, the so-called quantum nonlinearity parameter $\chi_0$ \cite{Mitter_1975, Ritus_1985, Ehlotzky_2009, Reiss_2009, Di_Piazza_2012, Dunne_2014}. The quantum nonlinearity parameter identifies the effective field scale at which a quantum process occurs in units of the QED field scale $F_{cr}=m^2/|e|=1.3\times 10^{16}\;\text{V/cm}$ ($=4.4\times 10^{13}\;\text{G}$), also known as the ``critical'' field of QED. If $F_0^{\mu\nu}$ is a measure of the amplitude of the background electromagnetic field and if $p^{\mu}$ is, for example, the initial four-momentum of an electron entering the external field and initiating the quantum process, the quantum nonlinearity parameter is defined as $\chi_0=\sqrt{|(F_0^{\mu\nu}p_{\nu})^2|}/mF_{cr}$. This definition indicates that the effective field scale at which the process occurs is the field that the electron experiences in its rest frame, which is a Lorentz-invariant quantity. Available laser and electron accelerator technology already allows for exploring the so-called SFQED regime ($\chi_0\gtrsim 1$) by combining either conventional \cite{Bula_1996, Burke_1997} or laser-based \cite{Leemans_2014} multi-GeV electron accelerators with high-power optical lasers \cite{Yanovsky_2008, CoReLS, APOLLON_10P, ELI}. Classical nonlinear effects have been already observed in recent experiments in laser-electron collision \cite{Khrennikov_2015, Yan_2017} and indications of nonlinear quantum effects have been reported in Refs.~\cite{Cole_2018, Poder_2018} (see Ref. \cite{Wistisen_2018} for a recent experiment within the quantum regime involving ultra-high energy positrons and a crystal).

Early SFQED experiments \cite{Bula_1996,Burke_1997} employed picosecond optical laser pulses focused on an area of the order of approximately $60\;\text{$\mu$m$^2$}$. This explains why the experimental results could be reproduced by calculating the corresponding QED processes in the presence of a monochromatic plane wave. Nowadays, experiments such as those reported in Ref.~\cite{Yan_2017, Cole_2018, Poder_2018} are carried out with femtosecond laser pulses focused down to a few square wavelengths, and future experiments aiming at even higher intensities will possibly employ shorter and more tightly focused laser pulses. The new features of ultraintense laser pulses used in experiments call for more general theoretical tools, suitable for describing qualitatively and quantitatively the experimental results in electromagnetic fields of a more complex spacetime structure than a plane wave. In SFQED the theoretical bottle neck is represented by the possibility of solving analytically the Dirac equation in the external field because the resulting electron states are then employed in the framework of the Furry picture \cite{Furry_1951, Landau_b_4_1982, Fradkin_b_1991}. This is indeed possible in the case of a plane wave \cite{Volkov_1935, Landau_b_4_1982}. In Refs.~\cite{Di_Piazza_2014, Di_Piazza_2015, Di_Piazza_2016, Di_Piazza_2017b} approximated expressions of the electron states (and of the propagator) in a tightly focused laser beam have been found and applied to investigate the two basic SFQED processes: nonlinear Compton scattering and nonlinear Breit-Wheeler pair production. The findings in Ref.~\cite{Di_Piazza_2014, Di_Piazza_2015, Di_Piazza_2016, Di_Piazza_2017b} are based on the assumption that the initial energy of the electron is so large that the electron itself is only barely deflected by the focused laser field. 

Another important theoretical tool to study SFQED processes occurring in the presence of virtually arbitrary electromagnetic fields is the so-called local-constant-field approximation (LCFA) \cite{Reiss_1962, Ritus_1985, Baier_b_1998}. This approximation is based on two physical observations. First, in SFQED experiments ultrarelativistic charged particles are typically employed, and in their rest frame they experience nearly equal transverse (with respect to the particles velocity in the laboratory frame) and mutually perpendicular electric and magnetic fields \cite{Jackson_b_1975}. Second, in SFQED experiments laser beams are typically employed, which are characterized by $\xi_0\gg 1$. Now, generally speaking, at $\xi_0\gg 1$ the probabilities of the basic SFQED processes, nonlinear Compton scattering and nonlinear Breit-Wheeler pair production, are formed on a space region (formation length) much shorter than the laser wavelength \cite{Ritus_1985, Baier_b_1998}. This implies that the probability of such processes in an arbitrary plane wave can be expressed as the integral over the laser phase of the corresponding probability in a constant crossed field evaluated at the local value of the plane-wave field. In this way, interference effects between contributions to the same final state originating from different formation lengths are ignored \cite{Harvey_2015}. These can be included via a saddle-point evaluation of the amplitudes, which has been carried out both in nonlinear Compton scattering \cite{Mackenroth_2011} and in nonlinear Breit-Wheeler pair production \cite{Meuren_2016}. A comprehensive benchmarking of the LCFA against the exact quantum emission probability can be found in Ref.~\cite{Blackburn_2018}, where particular emphasis is also put on the assumption that the photon is emitted along the instantaneous electron's velocity (collinear emission).

Now, it turns out that the formation length of the basic SFQED processes does not only depend on the parameter $\xi_0$ but it also depends on $\chi_0$ \cite{Baier_b_1998, Dinu_2016} and on the energies of the particles involved in the process \cite{Blankenbecler_1996, Baier_1989, Khokonov_2002, Wistisen_2015, Di_Piazza_2018c}. In particular, in Ref.~\cite{Di_Piazza_2018c} we have investigated the dependence of the formation length of nonlinear Compton scattering  on the energy of the emitted photon. We have shown that the parameter controlling the validity of the LCFA in a plane wave is given by
\begin{equation}
\label{eta_LCFA}
\eta_{\text{LCFA}}=\frac{p_--k_-}{k_-}\frac{\chi_0}{\xi_0^3}
\end{equation}
where $p_-$ and $k_-$ are the light-cone energies of the incoming electron and of the emitted photon, and where it is tacitly assumed that $(p_--k_-)\chi_0\gg k_-$. The LCFA turned out to be applicable if $\eta_{\text{LCFA}}\ll 1$. As a consequence, it is clear that even when $\xi_0\gg 1$, the LCFA becomes inapplicable for sufficiently small photon light-cone energies. Indeed, we have found that the exact expression of the differential probability $dP/dk_-$ of nonlinear Compton scattering tends to a constant in the limit $k_-\to 0$ rather than diverging as $k^{-2/3}_-$ as predicted by the LCFA. Finally, we have put forward a scheme to implement an improved expression of the emission probability for the collision of an electron beam with a laser pulse of a given central frequency $\omega_0$ in numerical codes, which takes into account the correct behavior in the infrared region.

In Ref.~\cite{Baier_1989} the leading-order correction with respect to the field derivatives of the differential probability of nonlinear Compton scattering was first calculated within the quasiclassical approximation. However, the resulting expression is more suitable for the central part of the emission spectrum and it becomes inapplicable in the low-energy region. Indeed, as we have pointed out in Ref.~\cite{Di_Piazza_2018c}, the energy integral of the differential probability of nonlinear Compton scattering including this leading-order correction diverges. Another attempt to improve the LCFA starting from the above-mentioned leading-order correction has been recently proposed in Ref.~\cite{Ilderton_2018b} for the case of a laser pulse, which includes a thorough analysis of the features of the infrared part of the emission spectrum. \corr{The method developed in Ref.~\cite{Ilderton_2018b} has the virtue of being based on the systematic approach introduced in Ref.~\cite{Baier_1989}, although the extension to the infrared region unavoidably requires by-hand adjustments. This method is applicable to situations where the electromagnetic fields feature a well-defined oscillation frequency $\omega_0$, as in the method presented in Ref.~\cite{Di_Piazza_2018c}.} In Ref.~\cite{Alexandrov_2019} the LCFA is scrutinized in the context of pair production in strong electromagnetic fields by studying the momentum spectra of the produced particles.

All the analytical results in Ref.~\cite{Di_Piazza_2018c} have been obtained in the case of a background plane wave. Here, we generalize those results to the case of background fields of virtually arbitrary spacetime structure, in the sense that a new prescription is put forward, which relies only on local quantities. In particular, the present prescription does not involve the parameter $\xi_0$, which in turns contains the frequency of the field. Correspondingly, we propose a more general implementation of an improved expression of the nonlinear Compton scattering emission probability into numerical codes, such as particle-in-cell (PIC) codes, able to describe electromagnetic fields with arbitrary spacetime structure. Note that the algorithms employed in PIC codes are based on differential emission probabilities expressed in terms of the local value of the background electromagnetic field. This is certainly not the case for the infrared region of the emission spectrum, because relatively low-energy photons correspondingly have long formation lengths. Thus, the implementation of the lower-energy part of the photon spectrum in terms of local quantities is unavoidably partially phenomenological. In this respect, we also report a mathematically and physically consistent approach to include higher-order effects in the LCFA, equivalent to that first proposed in Ref.~\cite{Baier_1989} (see also \cite{Ilderton_2018b}). Furthermore, we have explicitly calculated the leading-order correction of the differential probability not only of nonlinear Compton scattering but also of nonlinear Breit-Wheeler pair production \cite{Reiss_1962, Nikishov_1964, Narozhny_2000, Roshchupkin_2001, Reiss_2009, Heinzl_2010b, Mueller_2011b, Titov_2012, Nousch_2012, Krajewska_2013b, Jansen_2013, Augustin_2014, Meuren_2015, Meuren_2016} in a plane wave. Finally, the angularly-resolved spectra of nonlinear Compton scattering and of nonlinear Breit-Wheeler pair production within the LCFA have been reported.

\section{Improved LCFA for SFQED codes}
\label{Sec_II}
First, we review some of the key concepts and results obtained in Ref.~\cite{Di_Piazza_2018c}, then we show how the above results can be extended to more general field configurations. We start our analysis by considering an electron with incoming four-momentum $p^{\mu}=(\varepsilon,\bm{p})$, with $\varepsilon=\sqrt{m^2+\bm{p}^2}$, which collides with a plane wave propagating along the $\bm{n}$ direction ($\bm{n}^2=1$). The plane wave is characterized by the four-vector potential $A^{\mu}(\phi)=(0,\bm{A}_{\perp}(\phi))$, where $\phi=(nx)=t-\bm{n}\cdot\bm{x}$ [i.e., $n^{\mu}=(1,\bm{n})$], and where $\bm{n}\cdot\bm{A}_{\perp}(\phi)=0$ and $\lim_{\phi\to\pm\infty}\bm{A}_{\perp}(\phi)=\bm{0}$  [i.e., the four-potential is chosen in the Lorentz gauge $\partial_{\mu}A^{\mu}(\phi)=0$]. Since the plane wave depends only on the variable $\phi$, it is clearly convenient to introduce the light-cone coordinates $T=(t+\bm{n}\cdot\bm{x})/2$, $\bm{x}_{\perp}=\bm{x}-(\bm{n}\cdot\bm{x})\bm{n}$, and, indeed, $\phi=t-\bm{n}\cdot\bm{x}$, as well as the light-cone components $v_+=(v^0+\bm{n}\cdot\bm{v})/2$, $\bm{v}_{\perp}=\bm{v}-(\bm{n}\cdot\bm{v})\bm{n}$, and $v_-=v^0-\bm{n}\cdot\bm{v}$ of an arbitrary four-vector $v^{\mu}=(v^0,\bm{v})$ (note that $T=x_+$ and $\phi=x_-$). Assuming that the emitted photon (outgoing electron) is characterized by a four-momentum $k^{\mu}=(\omega,\bm{k})$, with $\omega=|\bm{k}|$ [$p^{\prime\mu}=(\varepsilon',\bm{p}')$, with $\varepsilon'=\sqrt{m^2+\bm{p}^{\prime\,2}}$], the leading-order differential emission probability $dP/dk_-$ averaged (summed) over all initial (final) discrete quantum numbers can be written in the form \cite{Di_Piazza_2018c}
\begin{widetext}
\begin{equation}
\label{dP_dk_-}
\begin{split}
\frac{dP}{dk_-}=&-i\frac{\alpha}{2\pi}\frac{1}{p_-}\frac{1}{\eta_0}\int\frac{d\varphi d\varphi'}{\varphi-\varphi'+i0}\bigg\{1+\frac{p_-^2+p^{\prime\,2}_-}{4p_-p'_-}[\bm{\xi}_{\perp}(\varphi)-\bm{\xi}_{\perp}(\varphi')]^2\bigg\}\\
&\times\exp\left\langle i\frac{1}{2\eta_0}\frac{k_-}{p'_-}\left\{\varphi-\varphi'+\int_{\varphi'}^{\varphi}d\tilde{\varphi}\,\bm{\xi}^2_{\perp}(\tilde{\varphi})-\frac{1}{\varphi-\varphi'}\left[\int_{\varphi'}^{\varphi}d\tilde{\varphi}\,\bm{\xi}_{\perp}(\tilde{\varphi})\right]^2\right\}\right\rangle,
\end{split}
\end{equation}
\end{widetext}
where $\eta_0=(k_0p)/m^2=\chi_0/\xi_0$, with $k_0^{\mu}=\omega_0n^{\mu}$,  where $\varphi=\omega_0\phi$ ($\varphi'=\omega_0\phi'$), where $p'_-=p_--k_-$, and where $\bm{\xi}_{\perp}(\varphi)=e\bm{A}_{\perp}(\varphi)/m$. Equation (\ref{dP_dk_-}) has been derived in detail in Ref.~\cite{Di_Piazza_2018c}. Here, we would like to derive the asymptotic limit $\lim_{k_-\to 0}dP/dk_-$ in a different and simpler way than in Ref.~\cite{Di_Piazza_2018c}. In fact, we first notice that in the limit $k_-\to 0$ one can neglect the field-dependent terms in the exponential function in Eq.~(\ref{dP_dk_-}) under the physically reasonable assumption that the function $\bm{\xi}_{\perp}(\varphi)$ is square integrable. \corr{Indeed, the field-dependent terms inside the braces of the exponential function in Eq.~(\ref{dP_dk_-}) are bounded for a finite duration pulse, whereas the term $\varphi-\varphi'$ is unbounded and must be retained because it gives a finite contribution even for $k_-\to 0$.} Then, we exploit the identity
\begin{equation}
\int_{-\infty}^{\infty}dx\frac{e^{iax}}{x+i0}=0,
\end{equation}
for any $a>0$, which is easily proved by means of the residue theorem. In this way, we obtain that 
\begin{equation}
\label{dP_dk_-_0}
\begin{split}
&\frac{dP_0}{dk_-}=\lim_{k_-\to 0}\frac{dP}{dk_-}=i\frac{\alpha}{2\pi}\frac{1}{p_-}\frac{1}{\eta_0}\int\frac{d\varphi d\varphi'}{\varphi-\varphi'+i0}\\
&\qquad\qquad\times \bm{\xi}_{\perp}(\varphi)\cdot\bm{\xi}_{\perp}(\varphi')e^{i\frac{1}{2\eta_0}\frac{k_-}{p'_-}(\varphi-\varphi')}\\
&=\frac{\alpha}{2\pi}\frac{1}{p_-}\frac{1}{\eta_0}\int_0^{\infty}d\rho\int d\varphi d\varphi'\bm{\xi}_{\perp}(\varphi)\cdot\bm{\xi}_{\perp}(\varphi')e^{i\rho(\varphi-\varphi')}\\
&=\frac{\alpha}{2\pi}\frac{1}{p_-}\frac{1}{\eta_0}\int_0^{\infty}d\rho|\tilde{\bm{\xi}}_{\perp}(\rho)|^2=\frac{\alpha}{2}\frac{1}{p_-}\frac{1}{\eta_0}\int d\varphi\,\bm{\xi}^2_{\perp}(\varphi),
\end{split}
\end{equation}
where we have introduced the Fourier transform $\tilde{\bm{\xi}}_{\perp}(\rho)$ of $\bm{\xi}_{\perp}(\varphi)$ and we have used the Parseval identity. This asymptotic behavior is qualitatively different from that predicted by the LCFA, which features an integrable divergence as $k_-^{-2/3}$ (see, e.g., Ref. \cite{Baier_b_1998}). Note that the constant in Eq. (\ref{dP_dk_-_0}) represents a non-perturbative result, i.e. it cannot be obtained by expanding in terms of the field derivatives [see the discussion below Eq.~(\ref{condition2})].

Another result we found in Ref.~\cite{Di_Piazza_2018c} is a general expression of the local formation phase $\varphi_f$ of nonlinear Compton scattering, which can be written as
\begin{equation}
\label{phi_f}
\varphi_f=\frac{8}{|\bm{\xi}'_{\perp}(\varphi)|}\sinh\left(\frac{1}{3}\sinh^{-1}\left(\frac{3\pi}{4}\frac{p'_-}{k _-}\chi(\varphi)\right)\right),
\end{equation}
where the prime in the symbol of a function indicates the derivative of the function with respect to the argument and where $\chi(\varphi)=\eta_0|\bm{\xi}'_{\perp}(\varphi)|$. \corr{For the sake of definiteness, we have chosen $2\pi$ as the typical phase where the plane wave changes substantially, such that the LCFA cannot be applied for $\varphi_f\gtrsim 2\pi$. In Ref.~\cite{Di_Piazza_2018c} we have also introduced the emitted photon light-cone energy $k_{-,\text{LCFA}}$ below which the LCFA is inapplicable. In fact, by imposing that at $k_-=k_{-,\text{LCFA}}$ the formation phase equals $2\pi$, we have obtained:}
\begin{equation}
\label{k_-_LCFA}
k_{-,\text{LCFA}}=\frac{p_-}{1+\frac{4}{3\pi\chi(\varphi)}\sinh\left(3\sinh^{-1}\left(\frac{\pi}{4}|\bm{\xi}'_{\perp}(\varphi)|\right)\right)},
\end{equation}
where we have pointed out that the quantity $k_{-,\text{LCFA}}(\varphi)$ depends on the laser phase. As we have shown in Ref.~\cite{Di_Piazza_2018c}, it is possible to derive the constant in Eq.~(\ref{dP_dk_-_0}) by starting from the cross section of linear Compton scattering. This led us to implement an improved expression of the probability $dP/dk_-$ with respect to the LCFA, which coincides with the latter for $k_-> k_{-,\text{LCFA}}(\varphi)$ and coincides with the probability obtained by starting from the cross section of linear Compton scattering for $k_-< k_{-,\text{LCFA}}(\varphi)$. Since for the light-cone energies of interest $k_{-,\text{LCFA}}(\varphi)\ll p_-$, we can further simplify the implementation put forward in Ref.~\cite{Di_Piazza_2018c} and directly use the constant in Eq.~(\ref{dP_dk_-_0}) for $k_-\le k_{-,\text{LCFA}}(\varphi)$. We have ensured that the numerical results obtained in Ref.~\cite{Di_Piazza_2018c} are only slightly affected by this modification, as it is also indicated by the fact that the exact probability is indeed practically constant for $k_-\le k_{-,\text{LCFA}}(\varphi)$ (see the numerical examples below and in Ref.~\cite{Di_Piazza_2018c}).

The \corr{above-mentioned} extension still uses the constant in Eq. (\ref{dP_dk_-_0}), which is intrinsically non-local as it contains the parameter $\xi_0$. In order to extend the method introduced in Ref.~\cite{Di_Piazza_2018c} to arbitrary field configurations, we have to rely only on local quantities. The starting point here is the following observation. As we have seen in Ref.~\cite{Di_Piazza_2018c} and as it can be seen from Eq. (\ref{k_-_LCFA}), we have that, in order of magnitude, $k_{-,\text{LCFA}}(\varphi)\sim p_-\chi_0/\xi_0^3$ for $\chi_0\lesssim 1$ and $\xi_0\gg 1$. This can also be equivalently obtained starting from Eq. (\ref{phi_f}) for $k_-\ll\min\{p_-,\chi_0p_-\}$. Now, the expression of the constant $dP_0/dk_-$ suggests to introduce a corresponding ``probability per unit phase''
\begin{equation}
\label{prob_lowk}
\frac{dP_0}{dk_-d\varphi}=\frac{\alpha}{2}\frac{1}{p_-}\frac{1}{\eta_0}\bm{\xi}^2_{\perp}(\varphi).
\end{equation}
Since for $k_-<k_{-,\text{LCFA}}(\varphi)$ the emission probability is practically constant, as a check of consistency, if we equate the differential probability $dP_0/dk_-d\varphi$ with the LCFA differential probability, we should find that the value of $k_-$ at which the two differential probabilities coincide is of the order of $k_{-,\text{LCFA}}(\varphi)$. This can indeed be easily proved by comparing the \corr{estimate}
\begin{equation}
\label{prob_lowk_appr}
\frac{dP_0}{dk_-d\varphi}\sim\frac{\alpha}{2}\frac{1}{p_-}\frac{1}{\eta_0}\xi_0^2
\end{equation}
with the differential emission probability within the LCFA in the region $k_-\ll \min\{p_-,\chi_0p_-\}$, which can be estimated as [see, e.g., Ref. \cite{Baier_b_1998} or Eqs. (\ref{dP_dk_-_LCFA}) and (\ref{K_nu_small_x}) below]
\begin{equation}
\label{prob_lcfa_lowk_appr}
\frac{dP_{\text{LCFA}}}{dk_-d\varphi}\sim\frac{\alpha}{\sqrt{3}\pi}\frac{\Gamma(2/3)3^{2/3}}{p_-\eta_0}\left(\frac{p_-\chi_0}{k_-}\right)^{2/3},
\end{equation}
where $\Gamma$ indicates the gamma function \cite{NIST_b_2010}. Now, by equating the right hand side of Eq.~(\ref{prob_lowk_appr}) and Eq.~(\ref{prob_lcfa_lowk_appr}), \corr{one can easily see that indeed the resulting value of $k_-$ is given by $k_-^*=3[2\Gamma(2/3)/\sqrt{3}\pi]^{3/2}p_-\chi_0/\xi_0^3\approx 1.05\,p_-\chi_0/\xi_0^3$. Since $k^*_-$ features the same parametric scaling of $k_{-,\text{LCFA}}(\varphi)$ at $k_-\ll\min\{p_-,\chi_0p_-\}$, this analysis} suggests to implement the improved version of the LCFA in the following straightforward way: use the differential probability within the LCFA for $k_->k_{-,\text{LCFA}}(\varphi)$ and use the constant value of this probability at $k_-=k_{-,\text{LCFA}}(\varphi)$ also for $k_-\le k_{-,\text{LCFA}}(\varphi)$. \corr{Moreover, if $k_{-,\text{LCFA}}(\varphi)$ coincided with $k_-^*$, one would reasonably expect that the resulting spectrum would feature the correct constant at low light-cone energies. In this respect, we notice that in the corresponding limiting region $k_-\ll\min\{p_-,\chi_0p_-\}$, one obtains starting from Eq. (\ref{phi_f}) that $k_{-,\text{LCFA}}(\varphi)\sim (12/\pi^2)p_-\chi_0/\xi_0^3\approx 1.2\,p_-\chi_0/\xi_0^3$, which suggests to preferably employ a matching point $k^*_{-,\text{LCFA}}(\varphi)=\rho_{\text{ft}}k_{-,\text{LCFA}}(\varphi)$, with $\rho_{\text{ft}}$ being a fine-tuning constant slightly smaller than unity (see below for additional details).}

The straightforward procedure described above for improving the LCFA can be generalized to a virtually arbitrary field if one provides a reliable definition of $k_{-,\text{LCFA}}(\varphi)$ for this more general case, which is also the case usually considered in numerical codes. Now, numerical codes do not work with light-cone energies but rather with the local values of the electron energy $\varepsilon(t)$ and with the emitted photon energy $\omega$. We have discussed in detail in Ref.~\cite{Di_Piazza_2018c} the relation among \corr{these} quantities and we have concluded that we can approximately replace $k_-$ with $2\omega$ and $p_-$ with the local value $2\varepsilon(t)$ in the physical regimes of interest. Indeed, discrepancies are expected to appear at such large emitted photon wavelengths that are resolved, e.g. by PIC codes, such that the effects of these photons are already taken into account via the numerical integration of the Maxwell-Lorentz system of equations. Correspondingly, in the case of a general background field, it is convenient to introduce a formation time $t_f$, which has to be compared with the time scale $\tau(t)$ where the background field changes locally. At first sight, one would expect to obtain $\tau(t)$ from the first time derivative of the Lorentz force. However, as discussed below [see Eqs.~(\ref{dP_dk_-_NLO}) and (\ref{dP_dk_-_NLO_s_km})], the leading-order correction to the LCFA in a plane wave depends not only on the first but also on the second time derivative of the fields, i.e., the contribution of the second derivative is of the same order of the contribution of the first derivative. Moreover, since the leading-order correction to the LCFA depends more precisely on the derivatives of the transverse Lorentz force $\bm{F}_{L,\perp}(t)$ with respect to the instantaneous electron velocity through the quantity $\chi(\varphi)$ [see Eqs.~(\ref{dP_dk_-_NLO}) and (\ref{dP_dk_-_NLO_s_km}) and Eq.~(\ref{chi_p})], it is natural to define $\tau(t)$ as
\begin{equation}
\label{tau}
\tau(t)=2\pi\sqrt{\frac{\bm{F}_{L,\perp}^2(t)}{\dot{\bm{F}}_{L,\perp}^2(t)+|\bm{F}_{L,\perp}(t)\cdot\ddot{\bm{F}}_{L,\perp}(t)|}},
\end{equation}
where the dots indicate the time derivative along the electron trajectory $\bm{x}(t)$. Three observations are in order: 1) another reason to employ the transverse Lorentz force $\bm{F}_{L,\perp}(t)$ is that in the ultrarelativistic limit the photon emission probability due to the longitudinal component of the Lorentz force is suppressed by a factor of the order of the square of the electron Lorentz factor $\gamma(t)=\varepsilon(t)/m$~\cite{Baier_b_1998}; 2) the constant $2\pi$ has been introduced in the definition of $\tau(t)$ such that in the collision with a monochromatic plane wave with amplitude $E_0$ and angular frequency $\omega_0$, one obtains $\tau(t)=\pi/\omega_0$ around the peaks of the field amplitude, where the improved scheme should basically coincide with the LCFA prediction for $\xi_0\gg 1$; and 3) the expression of $\tau(t)$ can be in principle employed as a starting point for the definition of a local field frequency and then of a local parameter $\xi_0$. As we have already noticed above, however, the present method does not require to introduce a local parameter $\xi_0$.

Having in mind the expression in Eq.~(\ref{phi_f}), we define the local formation time $t_f$ as 
\begin{equation}
\label{t_f}
\begin{split}
t_f(t)&=\frac{8\gamma(t)}{\chi(t)}\tau_C\sinh\left(\frac{1}{3}\sinh^{-1}\left(\frac{3\pi}{4}\frac{\varepsilon'(t)}{\omega}\chi(t)\right)\right),
\end{split}
\end{equation}
where $\tau_C=1/m\approx1.3\times 10^{-21}\;\text{s}$ is the Compton wavelength divided by the speed of light, $\varepsilon'(t)=\varepsilon(t)-\omega$ [it is clear that in this case the prime in $\varepsilon'(t)$ does not indicate a time derivative], and where $\chi(t)$ is the value of the quantum nonlinearity parameter along the electron trajectory.

At this point, the time-dependent threshold energy $\omega_{\text{LCFA}}(t)$ is defined as the value at which $t_f(t)=\tau(t)$:
\begin{equation}
\label{omegaLCFA}
\omega_{\text{LCFA}}(t)=\frac{\varepsilon(t)}{1+\frac{4}{3\pi\chi(t)}\sinh\left(3\sinh^{-1}\left(\frac{\chi(t)}{8\gamma(t)}\frac{\tau(t)}{\tau_C}\right)\right)}.
\end{equation}
Having defined $\tau(t)$ as in Eq.~(\ref{tau}), it follows that in the collision with a plane wave the definition of $\omega_{\text{LCFA}}(t)$ is in agreement with the definition of $k_{-,\text{LCFA}}(\varphi)$ in Eq. (\ref{k_-_LCFA}).

In conclusion, the above improved expression of the differential emission probability at a time $t$ is constructed in the following way:
\begin{enumerate}
\item use the expression $dP_{\text{LCFA}}(\omega,t)/d\omega dt$ of the emission probability within the LCFA for $\omega>\omega^*_{\text{LCFA}}(t)$, with $\omega^*_{\text{LCFA}}(t)=\rho_{\text{ft}}\omega_{\text{LCFA}}(t)$ [see the discussion below Eq. (\ref{prob_lcfa_lowk_appr})];
\item use the local constant value $dP_{\text{LCFA}}(\omega^*_{\text{LCFA}}(t),t)/d\omega dt$ for $\omega\le\omega^*_{\text{LCFA}}(t)$.
\end{enumerate}
\corr{In the next section, by comparing numerically the results obtained with this method to the full QED calculations, we fix the value of $\rho_{\text{ft}}$ to $\rho_{\text{ft}}=0.7$ in all simulations, which is indeed smaller than unity as expected from the discussion below Eq. (\ref{prob_lcfa_lowk_appr}).} We stress that the improved expression of the emission probability is applicable also for background fields of complex spacetime structure because it relies only on local quantities.

\section{Numerical implementation}

The implementation of the improved LCFA in SFQED codes based on a Monte Carlo approach is straightforward. The main difference with respect to the standard LCFA is that, at each timestep $\Delta t$, the code must calculate the local value of $\tau(t)$ as defined in Eq.~(\ref{tau}). Thus, in order to calculate $\dot{\bm{F}}_{L,\perp}(t)$ and $\ddot{\bm{F}}_{L,\perp}(t)$ numerically, for each particle the value of $\bm{F}_{L,\perp}(t)$ obtained at the previous two timesteps must be stored in addition to the local particle position $\bm{x}(t)$ and momentum $\bm{p}(t)$. In addition, a Boolean variable that takes into account whether the particle is new, i.e. it has just been created, is also stored for each particle. This is needed because electrons and positrons can be created during the simulation due to, e.g., nonlinear Breit-Wheeler pair production, and for new particles the value of $\bm{F}_{L,\perp}(t)$ is not available at previous timesteps, such that the standard LCFA should be employed for these particles, initially. 

In the following, we report an explicit recipe for the implementation of the improved LCFA with leapfrog integration \cite{Birdsall_b_2004}, which is widely employed in PIC codes. In order to take into account the effect of the recoil on the particle motion and to minimize the changes in the existing codes, we employ the result obtained in Ref.~\cite{Tamburini_2010}. Following Ref.~\cite{Tamburini_2010}, starting from $\bm{p}^{(n-1/2)}$, the total momentum after one step $\bm{p}^{(n+1/2)}$ of a particle under the influence both of the Lorentz $\bm{F}_{L}(t)$ and of an ``effective'' recoil force $\bm{F}_{R}(t)$ can be obtained as \corr{$\bm{p}^{(n+1/2)} = \bm{p}_{L}^{(n+1/2)} + \int_{t^{(n)}-\Delta t/2}^{t^{(n)}+\Delta t/2} dt\,\bm{F}_{R}^{(n)}$}, where $\bm{p}_{L}^{(n+1/2)}$ is the momentum obtained by advancing $\bm{p}^{(n-1/2)}$ of one step by using the already existing leapfrog integrator for the Lorentz force, and
\corr{\begin{equation}
\bm{F}_{R}^{(n)} = 
\begin{cases}
0 & \text{no emission}\\
- \delta(t-t^{(n)}) \omega \bm{p}^{(n)}/|\bm{p}^{(n)}| & \text{emission}
\end{cases}
\end{equation}
i}s the ``effective'' recoil force. Thus, the algorithm is as follows: (1)~starting from $\bm{x}^{(n)}$, $\bm{p}^{(n-1/2)}$ calculate $\bm{p}_{L}^{(n+1/2)}$ with the leapfrog integrator existing in the code; (2)~calculate:
\begin{align}
\label{fl}
\bm{F}_{L}^{(n)} & = \frac{\bm{p}_{L}^{(n+1/2)}-\bm{p}^{(n-1/2)}}{\Delta t}, \\
\bm{p}_{L}^{(n)} & = \frac{\bm{p}_{L}^{(n+1/2)}+\bm{p}^{(n-1/2)}}{2}, \\
\gamma^{(n)} & = \sqrt{1+(\bm{p}_{L}^{(n)}/m)^2}, \\
\label{fl_perp}
\bm{F}_{L,\perp}^{(n)} & = \bm{F}_{L}^{(n)} - \frac{(\bm{F}_{L}^{(n)} \cdot \bm{p}_{L}^{(n)})}{(m \gamma^{(n)})^2} \bm{p}_{L}^{(n)}, \\
\label{chi_p}
\chi^{(n)} & = \tau_C \gamma^{(n)} \sqrt{\left(\frac{\bm{F}_{L,\perp}^{(n)}}{m}\right)^2}.
\end{align}
Concerning Eq. (\ref{fl}), we remind that the leapfrog integrator does not evaluate $\bm{F}_{L}^{(n)}$ explicitly for computing $\bm{p}_{L}^{(n+1/2)}$ from $\bm{p}^{(n-1/2)}$ \cite{Birdsall_b_2004}. Note that, to avoid possible issues for $|\bm{p}_{L}^{(n)}|\approx 0$, in Eqs.~(\ref{fl_perp})-(\ref{chi_p}) we have used the approximation $\bm{p}_{L}^{(n)} / |\bm{p}_{L}^{(n)}| \approx \bm{p}_{L}^{(n)} / m \gamma^{(n)}$, which is an excellent approximation already for $\gamma^{(n)} \gtrsim 10$. We stress that the approximation of collinear emission and the equations for the standard LCFA are anyway applicable only for $\gamma(t)\gg1$. Now, if the particle is new, set $\bm{F}_{L,\perp}^{(n-2)}=\bm{F}_{L,\perp}^{(n-1)}=\bm{F}_{L,\perp}^{(n)}$ and change its Boolean variable to false. As it will be clear in the following, this implies that the particle will emit photons according to the standard LCFA, initially (note that, since the emission probability in each time step $\Delta t$ needs to be much smaller than unity, this assumption has no sizable effects); (3)~calculate
\begin{align}
\dot{\bm{F}}_{L,\perp}^{(n)} & = \frac{\bm{F}_{L,\perp}^{(n)} - \bm{F}_{L,\perp}^{(n-1)}}{\Delta t}, \\
\ddot{\bm{F}}_{L,\perp}^{(n)} & = \frac{\bm{F}_{L,\perp}^{(n)} - 2 \bm{F}_{L,\perp}^{(n-1)} + \bm{F}_{L,\perp}^{(n-2)}}{(\Delta t)^2}, \\
\delta^{(n)} = & \tau_C^2 \left[ (\dot{\bm{F}}_{L,\perp}^{(n)})^2 + |\bm{F}_{L,\perp}^{(n)} \cdot \ddot{\bm{F}}_{L,\perp}^{(n)}| \right];
\end{align}
(4)~if $(\gamma^{(n)})^2 \delta^{(n)}/\zeta^2 > (\chi^{(n)})^2 (\bm{F}_{L,\perp}^{(n)})^2$, where $\zeta$ is a nearly negligible number relative to unity (with, e.g., double precision $\zeta \approx 2.22 \times 10^{-16}$), from Eq.~(\ref{tau}) calculate 
\begin{equation}
\frac{\tau^{(n)}}{\tau_C} = 2 \pi \sqrt{\frac{(\bm{F}_{L,\perp}^{(n)})^2}{\delta^{(n)}}},
\end{equation}
otherwise the background fields are basically constant, and the LCFA applies throughout the photon spectrum. This condition is introduced to avoid numerical issues for constant background fields, where the LCFA holds and $\tau^{(n)}/\tau_C$ diverges; (5)~from Eq.~(\ref{omegaLCFA}), if $\chi^{(n)}>\zeta$ calculate $\corr{\omega_{\text{LCFA}}^{*(n)}=\rho_{\text{ft}}\omega_{\text{LCFA}}^{(n)}}$, with
\begin{equation}
\label{omegaLCFA_n}
\omega_{\text{LCFA}}^{(n)} = \frac{\varepsilon^{(n)}}{1 + \frac{4}{3\pi \chi^{(n)}}\sinh\left(3\sinh^{-1}\left(\frac{\chi^{(n)}}{8 \gamma^{(n)}}\frac{\tau^{(n)}}{\tau_C}\right)\right)}.
\end{equation}
Note that the condition $\chi^{(n)}>\zeta$ is introduced to avoid possible numerical issues, but has no practical effect as the probability of emission per unit time is proportional to $\chi$ and the emitted power scales as $\chi^2$, such that emissions for $\chi^{(n)}\leq\zeta$ are completely negligible. In addition, if $(\varepsilon^{(n)} - \omega_{\text{LCFA}}^{*(n)}) < 10^{-3} \varepsilon^{(n)}$ no emission is deemed at this timestep. This second condition is of physical origin, as it implies that $\omega_{\text{LCFA}}^{*(n)} \approx \varepsilon^{(n)}$, such that the LCFA basically cannot be applied throughout the \corr{entire} photon spectrum. Note that these are very rare processes almost exclusively occurring at very low field amplitudes, and can be neglected; (6)~following the physical argument below Eq.~(\ref{prob_lowk}), evaluate $d P_{\text{LCFA}}(\omega,t)/d\omega dt$ at $\omega_{\text{LCFA}}^{*(n)}$. The improved emission spectrum is equal to a constant with value $d P_{\text{LCFA}}(\omega_{\text{LCFA}}^{*(n)},t)/d\omega dt$ for $\omega < \omega_{\text{LCFA}}^{*(n)}$, while it is equal to the standard LCFA spectrum for $\omega > \omega_{\text{LCFA}}^{*(n)}$. Now, the rate of photon emission per unit time is simply the sum of $\omega_{\text{LCFA}}^{*(n)} d P_{\text{LCFA}}(\omega_{\text{LCFA}}^{*(n)},t)/d\omega dt$ and $\int_{\omega_{\text{LCFA}}^{*(n)}}^{\varepsilon^{(n)}}{d\omega \, [d P_{\text{LCFA}}(\omega,t)/d\omega dt]}$, and the algorithm of the Monte Carlo method for determining whether a photon emission occurs and, if deemed, the energy of the emitted photon follows the same steps as for the standard LCFA (see, e.g., the Supplementary information of Ref.~\cite{Tamburini_2017}); (7)~following the argument above Eq.~(\ref{fl}), if a photon emission occurs the electron momentum becomes: $\bm{p}^{(n+1/2)} = \bm{p}_{L}^{(n+1/2)} - \omega \bm{p}_{L}^{(n)} / |\bm{p}_{L}^{(n)}| \approx \bm{p}_{L}^{(n+1/2)} - \omega  \bm{p}_{L}^{(n)} / m \gamma^{(n)}$; \corr{(8)~advance the particle position $\bm{x}^{(n+1)} = \bm{x}^{(n)} + \bm{p}^{(n+1/2)} \Delta t/\gamma^{(n+1/2)} m$, where $\gamma^{(n+1/2)} = \sqrt{1 + (\bm{p}^{(n+1/2)}/m)^2}$, and store the value of $\bm{F}_{L,\perp}^{(n-1)}$, $\bm{F}_{L,\perp}^{(n)}$, and of the Boolean variable together with $\bm{x}^{(n+1)}$, $\bm{p}^{(n+1/2)}$.}

\begin{widetext}
\begin{figure}
\begin{center}
\includegraphics[width=\columnwidth]{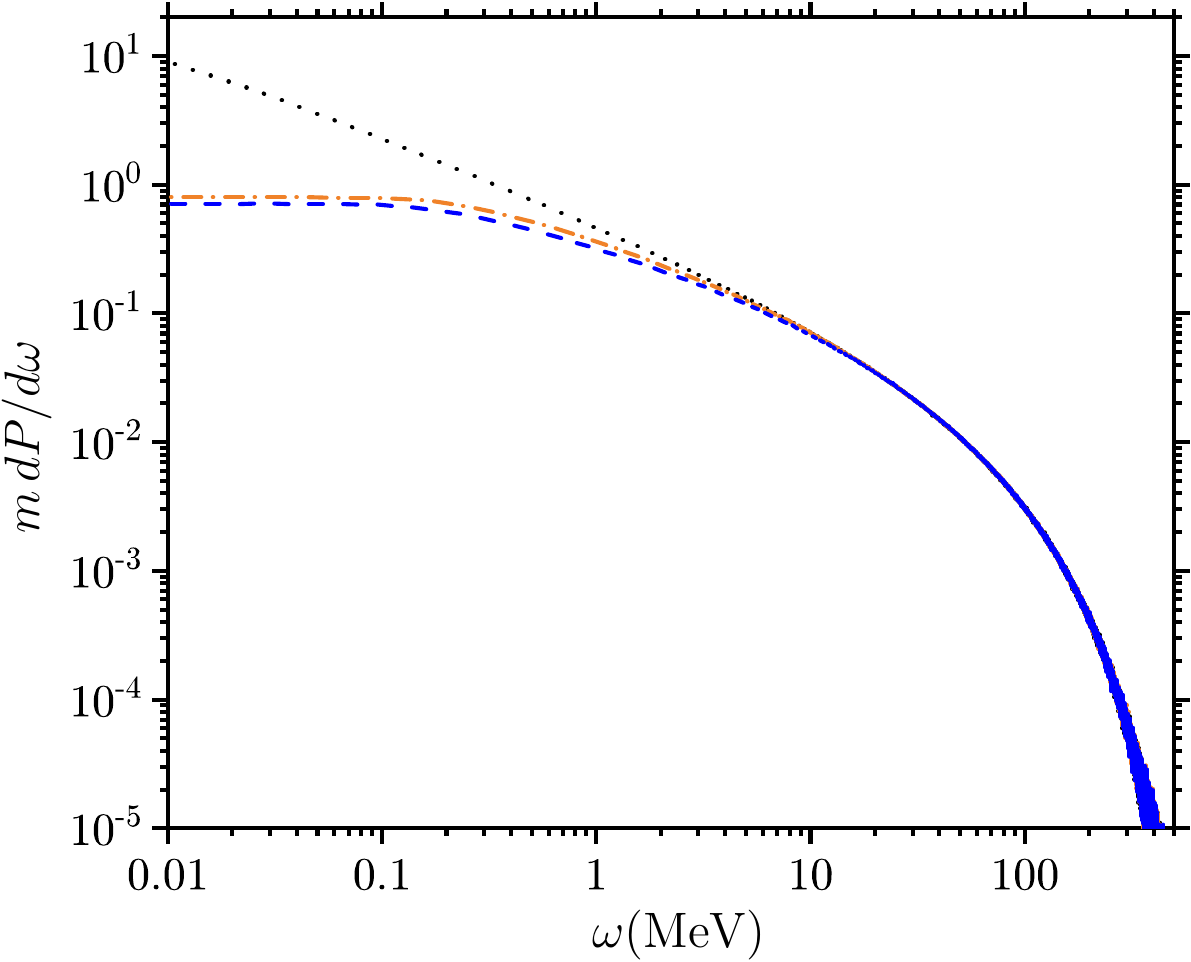}
\caption{Local-constant-field-approximated differential photon emission probability (dotted black curve) compared to the improved probability presented in Ref.~\cite{Di_Piazza_2018c} (dash-dotted orange curve) and the one developed here (dashed blue curve) for a beam of $10^8$ electrons with Gaussian distribution in space and in momentum, 3~$\mu$m diameter (FWHM), 13~$\mu$m length FWHM, 1~GeV mean energy, 100~MeV energy width (FWHM), and 1~mrad angular aperture colliding head-on with a laser pulse with 45~fs duration (FWHM of the intensity), 4~$\mu$m waist radius, and $4.4\times10^{20}$~W/cm$^2$ peak intensity ($\xi_0\approx10$).} 
\label{fig:3}
\end{center}
\end{figure}
\end{widetext}

\begin{widetext}

\subsection{Numerical examples}

In this section we report the results of numerical simulations carried out with the prescription described above. \corr{As we have already mentioned, in all numerical simulations we employed the same value of the fine-tuning constant, which was set to $\rho_{\text{ft}}=0.7$. This value of the constant has provided the best agreement between the simulation results and the full QED calculations in all examples reported below (see Figs.~\ref{fig:1}-\ref{fig:2}).}

In the first two examples, shown in Figs.~\ref{fig:1}a and \ref{fig:1}b, we \corr{report the simulation results with the same parameters as those in Ref.~\cite{Di_Piazza_2018c}} in order to compare the two methods \corr{(compare in particular the dash-dotted orange curve, corresponding to the method presented in Ref.~\cite{Di_Piazza_2018c}, with the dashed blue curve, corresponding to the present method). The agreement between the two methods is very good, with the present approach being even superior at intermediate energies and having the crucial} advantages that it is applicable to arbitrary and complex background fields and that it can be easily implemented in PIC codes, as we have discussed in detail in the previous section.
\begin{figure}
\begin{center}
\includegraphics[width=\columnwidth]{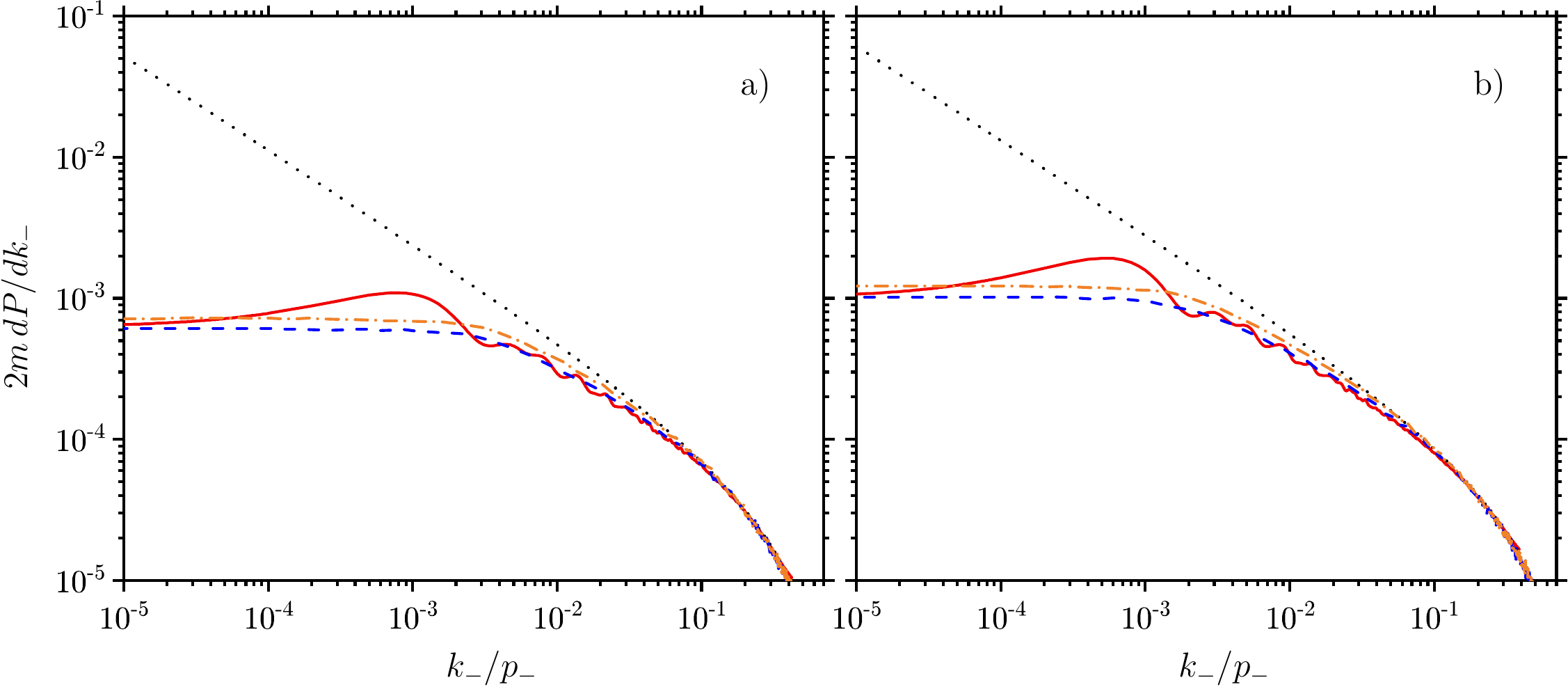}
\caption{Exact (solid red curve) vs local-constant-field-approximated (dotted black curve) differential photon emission probability for an electron with initial energy of $10\;\text{GeV}$ colliding head-on with a plane wave pulse of $5\;\text{fs}$ duration (full width at half maximum (FWHM) of the intensity), and $2.7\times 10^{20}\;\text{W/cm$^2$}$ peak intensity ($\xi_0\approx 8$) for panel a) and $4.4\times 10^{20}\;\text{W/cm$^2$}$ peak intensity ($\xi_0\approx 10$) for panel b). The dashed blue curve (dash-dotted orange curve) shows the same differential probability obtained via the numerical code presented in Ref.~\cite{Tamburini_2017}, with the improved emission model as described in the text (with the model presented in Ref.~\cite{Di_Piazza_2018c}).} \label{fig:1}
\end{center}
\end{figure}
\corr{In Fig.~\ref{fig:2} the results of other simulations are presented with different numerical parameters. In particular, we have chosen a peak laser intensity of $4\times 10^{19}\;\text{W/cm$^2$}$ ($\xi_0\approx 3$) in  Fig.~\ref{fig:2}a, a laser pulse duration of $10\;\text{fs}$ in Fig.~\ref{fig:2}b, and an initial electron energy of $5\;\text{GeV}$ in  Fig.~\ref{fig:2}c. The very good agreement between our improved method and the exact quantum calculations shows the broad applicability and the robustness of the employed numerical value of $\rho_{\text{ft}}$.}
\begin{figure}
\begin{center}
\includegraphics[width=\columnwidth]{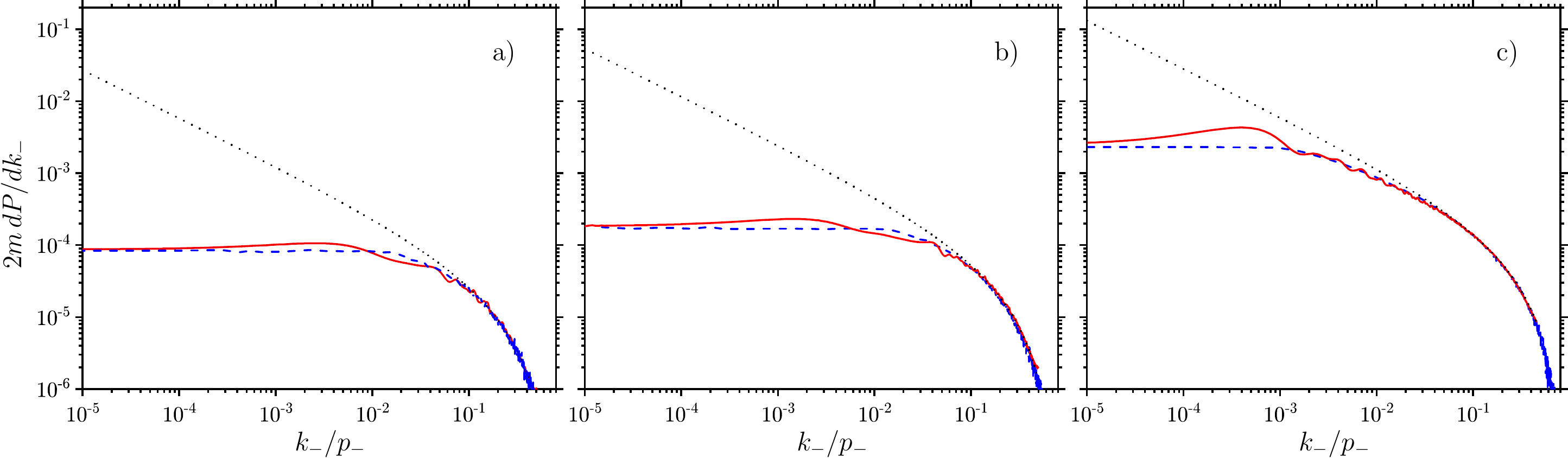}
\caption{\corr{Exact (solid red curve) vs local-constant-field-approximated (dotted black curve) differential photon emission probability for the following numerical parameters. Panel a): electron initial energy of $10\;\text{GeV}$, plane-wave pulse duration of $5\;\text{fs}$ (FWHM of the intensity), and of peak intensity $4\times 10^{19}\;\text{W/cm$^2$}$ ($\xi_0\approx 3$); Panel b): electron initial energy of $10\;\text{GeV}$, plane-wave pulse duration of $10\;\text{fs}$ (FWHM of the intensity), and of peak intensity $4\times 10^{19}\;\text{W/cm$^2$}$ ($\xi_0\approx 3$); Panel c): electron initial energy of $5\;\text{GeV}$, plane-wave pulse duration of $5\;\text{fs}$ (FWHM of the intensity), and of peak intensity $2.7\times 10^{20}\;\text{W/cm$^2$}$ ($\xi_0\approx 8$). The dashed blue curve shows the same probability obtained via the numerical code presented in Ref.~\cite{Tamburini_2017}, with the improved emission model as described in the text.}} \label{fig:2}
\end{center}
\end{figure}
\end{widetext}
Finally, Fig.~\ref{fig:3} displays the results obtained from the more realistic simulation of a beam of $10^8$ electrons with Gaussian distribution in space and in momentum, 3~$\mu$m diameter FWHM, 13~$\mu$m length FWHM, 1~GeV mean energy, 100~MeV energy width (FWHM), and 1~mrad angular aperture colliding head-on with a laser pulse with 45~fs duration (FWHM of the intensity), 4~$\mu$m waist radius, and $4.4\times10^{20}$~W/cm$^2$ peak intensity ($\xi_0\approx10$). Figure~\ref{fig:3} reports the results obtained with the standard LCFA model (dotted black curve), the model introduced in Ref.~\cite{Di_Piazza_2018c} (dash-dotted orange curve), and the new model presented here (dashed blue curve). \corr{Note that since the average electrons beam energy is much larger than $m\xi_0$, the electrons are barely deviated from their initial propagation direction such that they only weakly experience the transverse structure of the laser pulse, and the method proposed in Ref.~\cite{Di_Piazza_2018c} is applicable (also because the background field has a well-defined central frequency $\omega_0$).} The important message of \corr{Fig.~\ref{fig:3}, given the realistic parameters employed,} is that the failure of the LCFA at low frequencies is potentially observable experimentally with currently existing lasers in a compact, all-optical setup. For example, the LCFA predicts that a total number of about $8.6\times 10^8$ photons are emitted, whereas the two improved models predict $6.6\times 10^8$ (the one presented in Ref.~\cite{Di_Piazza_2018c}) and $6.2\times 10^8$ (the one presented here).
\section{Systematic approach to high-order corrections to the LCFA}

In this section we provide a systematic approach to evaluate the differential probability of nonlinear Compton scattering and of nonlinear Breit-Wheeler pair production in a plane wave. For this reason, it is convenient to add the index $C$ to the probabilities investigated in the previous sections and to indicate as $dP_{BW}/dp_-$, the corresponding differential probability of nonlinear Breit-Wheeler pair production per unit of produced positron light-cone energy (the reason to use again the symbol $p_-$ will be clear below). Our starting point is to write the differential probability $dP_C/dk_-$ in the form $dP_C/dk_-=\int d\varphi_+\,dP_C/dk_-d\varphi_+$, where [see Eqs. (11)-(12) in Ref.~\cite{Di_Piazza_2018c})
\begin{widetext}
\begin{gather}
\label{f}
\frac{dP_C}{dk_-d\varphi_+}=\frac{\alpha}{2\pi}\frac{1}{p_-}\frac{1}{\eta_0}\text{Im}\int\frac{d\varphi_-}{\varphi_-+i0}\left[1+\frac{p_-^2+p_-^{\prime\,2}}{4p_-p'_-}\bm{\Xi}^2_{\perp}(\varphi_-,\varphi_+)\right]e^{i\Phi_C(k_-,\varphi_-,\varphi_+)},\\
\label{Phi}
\Phi_C(k_-,\varphi_-,\varphi_+)=\frac{1}{2\eta_0}\frac{k_-}{p'_-}\Bigg\{\varphi_-+\int_{-\varphi_-/2}^{\varphi_-/2}d\tilde{\varphi}\,\bm{\xi}^2_{\perp}(\varphi_++\tilde{\varphi})-\frac{1}{\varphi_-}\bigg[\int_{-\varphi_-/2}^{\varphi_-/2}d\tilde{\varphi}\,\bm{\xi}_{\perp}(\varphi_++\tilde{\varphi})\bigg]^2\Bigg\}.
\end{gather}
\end{widetext}
Here, we have introduced the variables $\varphi_+=(\varphi+\varphi')/2$ and $\varphi_-=\varphi-\varphi'$ and the convenient quantity
\begin{equation}
\bm{\Xi}_{\perp}(\varphi_-,\varphi_+)=\bm{\xi}_{\perp}\left(\varphi_++\frac{\varphi_-}{2}\right)-\bm{\xi}_{\perp}\left(\varphi_+-\frac{\varphi_-}{2}\right).
\end{equation}
and we will closely follow the approach first outlined in Ref.~\cite{Baier_1989}. This will also give us the possibility of establishing quantitatively the limits of validity of the LCFA and the size of the expected corrections. In fact, we recall that the LCFA is applicable when the laser formation phase $\varphi_f$ is much smaller than $2\pi$, such that one can expand the field-dependent terms in Eqs. (\ref{f})-(\ref{Phi}) around $\varphi_+$ \corr{[more precisely the function $\bm{\Xi}_{\perp}(\varphi_-,\varphi_+)$ is expanded around the point $(0,\varphi_+)$]}. In particular, the (leading-order) LCFA corresponds to expand the function $\bm{\Xi}_{\perp}(\varphi_-,\varphi_+)$ up to first order in $\varphi_-$ and the phase $\Phi(k_-,\varphi_-,\varphi_+)$ up to the third order in $\varphi_-$:
\begin{align}
\bm{\Xi}_{\perp}(\varphi_-,\varphi_+)&\approx\bm{\xi}'_{\perp}(\varphi_+)\varphi_-\equiv\bm{\Xi}_{\perp,\text{LCFA}}(\varphi_-,\varphi_+),\\
\begin{split}
\Phi_C(k_-,\varphi_-,\varphi_+)&\approx\frac{1}{2\eta_0}\frac{k_-}{p'_-}\left[\varphi_-+\frac{\bm{\xi}^{\prime\,2}_{\perp}(\varphi_+)}{12}\varphi_-^3\right]\\
&\equiv\Phi_{C,\text{LCFA}}(k_-,\varphi_-,\varphi_+).
\end{split}
\end{align}
By employing these approximated quantities, one can carry out the integral over $\varphi_-$ (see the Appendix and Ref. \cite{Baier_b_1998}) and obtain the differential probability within the LCFA, which is given by 
\begin{equation}
\label{dP_dk_-_LCFA}
\begin{split}
\frac{dP_{C,\text{LCFA}}}{dk_-d\varphi_+}=&\frac{\alpha}{\sqrt{3}\pi}\frac{1}{p_-}\frac{1}{\eta_0}\left[\frac{p_-^2+p_-^{\prime\,2}}{p_-p'_-}\text{K}_{2/3}\left(\frac{2}{3}\frac{k_-}{p'_-}\frac{1}{\chi(\varphi_+)}\right)\right.\\
&\left.-\int_{2k_-/3p'_-\chi(\varphi_+)}^{\infty} dz\,\text{K}_{1/3}(z)\right],
\end{split}
\end{equation}
where $\text{K}_{\nu}$ indicates the modified Bessel function of order $\nu$ \cite{NIST_b_2010}. \corr{Note that, although we have employed the approximated expression of the integrand valid for small $\varphi_-$, the integration in $\varphi_-$ can be safely extended up to $\varphi_-\to\pm\infty$ because the contribution to the integral at large values of $|\varphi_-|$ is anyway negligible due to the fast oscillations of the integrand itself.}
In Ref.~\cite{Di_Piazza_2018c} we have already calculated the leading-order correction $\delta\Phi_C(k_-,\varphi_-,\varphi_+)$ to the phase $\Phi_{C,\text{LCFA}}(k_-,\varphi_-,\varphi_+)$, which is obtained by expanding $\Phi_C(k_-,\varphi_-,\varphi_+)$ up to terms of the order of $\varphi_-^5$ (see also Ref. \cite{Baier_1989,Khokonov_2002,Ilderton_2018b}):
\begin{equation}
\begin{split}
\delta\Phi_C(k_-,\varphi_-,\varphi_+)=&\frac{1}{2\eta^3_0}\frac{k_-}{p'_-}\\
&\times\frac{\bm{\chi}^{\prime\,2}(\varphi_+)+3\bm{\chi}(\varphi_+)\cdot\bm{\chi}^{\prime\prime}(\varphi_+)}{720}\varphi_-^5,
\end{split}
\end{equation}
where we have set $\bm{\chi}(\varphi)=\eta_0\bm{\xi}'_{\perp}(\varphi)$ (note that $\chi(\varphi)=|\bm{\chi}(\varphi)|$). Analogously, we obtain the leading-order correction $\delta\bm{\Xi}_{\perp}(\varphi_-,\varphi_+)$ to $\bm{\Xi}_{\perp,\text{LCFA}}(\varphi_-,\varphi_+)$, which is given by
\begin{equation}
\delta\bm{\Xi}_{\perp}(\varphi_-,\varphi_+)=\frac{1}{\eta_0}\frac{\bm{\chi}^{\prime\prime}(\varphi_+)}{24}\varphi_-^3
\end{equation}
After plugging the expressions of $\delta\Phi_C(k_-,\varphi_-,\varphi_+)$ and of $\delta\bm{\Xi}_{\perp}(\varphi_-,\varphi_+)$ in Eqs. (\ref{f})-(\ref{Phi}) and by employing the formulas in the Appendix, one easily obtains the photon emission differential probability up to next-to-leading order in the LCFA
\begin{widetext}
\begin{equation}
\label{dP_dk_-_NLO}
\begin{split}
&\frac{dP_C^{\text{NLO}}}{dk_-d\varphi_+}=\frac{dP_{C,\text{LCFA}}}{dk_-d\varphi_+}\\
&\quad+\frac{\alpha}{\sqrt{3}\pi}\frac{\eta_0}{p_-}\left\langle\frac{1}{45}\frac{\bm{\chi}^{\prime\,2}(\varphi_+)+3\bm{\chi}(\varphi_+)\cdot\bm{\chi}^{\prime\prime}(\varphi_+)}{\chi^4(\varphi_+)}\left[\frac{k_-}{p'_-}\frac{1}{\chi(\varphi_+)}\text{K}_{1/3}\left(\frac{2}{3}\frac{k_-}{p'_-}\frac{1}{\chi(\varphi_+)}\right)-2\text{K}_{2/3}\left(\frac{2}{3}\frac{k_-}{p'_-}\frac{1}{\chi(\varphi_+)}\right)\right]\right.\\
&+\frac{1}{45}\frac{\bm{\chi}^{\prime\,2}(\varphi_+)+3\bm{\chi}(\varphi_+)\cdot\bm{\chi}^{\prime\prime}(\varphi_+)}{\chi^4(\varphi_+)}\frac{p_-^2+p_-^{\prime\,2}}{p_-p'_-}\\
&\times\left\{6\text{K}_{2/3}\left(\frac{2}{3}\frac{k_-}{p'_-}\frac{1}{\chi(\varphi_+)}\right)-\left[\frac{k_-}{p'_-}\frac{1}{\chi(\varphi_+)}+4\frac{p'_-}{k_-}\chi(\varphi_+)\right]\text{K}_{1/3}\left(\frac{2}{3}\frac{k_-}{p'_-}\frac{1}{\chi(\varphi_+)}\right)\right\}\\
&\left.-\frac{1}{3}\frac{p_-^2+p_-^{\prime\,2}}{p_-p'_-}\frac{\bm{\chi}(\varphi_+)\cdot\bm{\chi}^{\prime\prime}(\varphi_+)}{\chi^4(\varphi_+)}\left[\text{K}_{2/3}\left(\frac{2}{3}\frac{k_-}{p'_-}\frac{1}{\chi(\varphi_+)}\right)-\frac{p'_-}{k_-}\chi(\varphi_+)\text{K}_{1/3}\left(\frac{2}{3}\frac{k_-}{p'_-}\frac{1}{\chi(\varphi_+)}\right)\right]\right\rangle.
\end{split}
\end{equation}
\end{widetext}

The expression of the differential emission probability in Eq. (\ref{dP_dk_-_NLO}) is in agreement with the results in Ref.~\cite{Baier_1989}, which can be obtained with the substitution rules $\eta_0\bm{\xi}'_{\perp}(\varphi)\to \varepsilon\bm{b}/m^2$ and $\eta_0 d/d\varphi\to\varepsilon m^{-2}\bm{V}\cdot\bm{\nabla}$, and suggests that the corrections to the LCFA scale as $1/\xi_0^2$ if $\chi_0\lesssim 1$ and $k_-\sim p_-$. This explains previous findings showing that the LCFA in nonlinear Breit-Wheeler pair production \cite{Meuren_2016} and nonlinear Compton scattering \cite{Blackburn_2018} turned out to be a satisfactory approximation at the percentage level already for $\xi_0\gtrsim 5$. However, if $k_-\ll p_-$, we have already seen in Ref.~\cite{Di_Piazza_2018c} that the LCFA may fail, i.e., the corrections can be large even if $\xi_0\gg 1$. In order to ascertain this explicitly, we can expand the expression in Eq. (\ref{dP_dk_-_NLO}) for $k_-/p_-\ll \min(1,\chi_0)$.

We recall that (see, e.g., Ref. \cite{NIST_b_2010})
\begin{align}
\label{K_nu_small_x}
\text{K}_{\nu}(x)\approx \frac{\Gamma(\nu)}{2}\Big(\frac{2}{x}\Big)^{\nu}, && 0<x\ll 1.
\end{align}
By employing this approximated expression it is easy to show that for $k_-/p_-\ll \min(1,\chi_0)$, it is
\begin{widetext}
\begin{equation}
\label{dP_dk_-_NLO_s_km}
\frac{dP_C^{\text{NLO}}}{dk_-d\varphi_+}\approx\frac{\alpha}{\sqrt{3}\pi}\frac{1}{p_-}\frac{1}{\eta_0}\Gamma\Big(\frac{2}{3}\Big)\left[3\frac{p_-}{k_-}\chi(\varphi_+)\right]^{2/3}\left\{1-\frac{\eta_0^2}{135}\frac{\Gamma(1/3)}{\Gamma(2/3)}\frac{4\bm{\chi}^{\prime\,2}(\varphi_+)-3\bm{\chi}(\varphi_+)\cdot\bm{\chi}^{\prime\prime}(\varphi_+)}{\chi^4(\varphi_+)}\left[3\frac{p_-}{k_-}\chi(\varphi_+)\right]^{2/3}\right\}.
\end{equation}
\end{widetext}
On the one hand this result indicates that the corrections are small if $\eta_{\text{LCFA}}\ll 1$ [see Eq. (\ref{eta_LCFA}) recalling that here it is $k_-/p_-\ll \min(1,\chi_0)$]. Indeed, the corrections scale as $\eta^{2/3}_{\text{LCFA}}$ in order of magnitude, in agreement with the results in Ref.~\cite{Di_Piazza_2018c}. More accurately there are two conditions of validity, which depend on the structure of the plane wave and which read
\begin{align}
\label{condition1}
\eta_0^2\frac{\bm{\chi}^{\prime\,2}(\varphi_+)}{\chi^4(\varphi_+)}\Big[\frac{p_-}{k_-}\chi(\varphi_+)\Big]^{2/3}\ll 1,\\
\label{condition2}
\eta_0^2\frac{|\bm{\chi}(\varphi_+)\cdot\bm{\chi}^{\prime\prime}(\varphi_+)|}{\chi^4(\varphi_+)}\Big[\frac{p_-}{k_-}\chi(\varphi_+)\Big]^{2/3}\ll 1.
\end{align}
On the other hand, as it has been observed in Ref.~\cite{Di_Piazza_2018c,Ilderton_2018b}, the leading-order correction induces a non-integrable divergence for $k_-\to 0$, which is in agreement with the fact that the approximation breaks down for too-low light-cone energies. This also implies that from the point of view of the expansion with respect to the field derivatives, the constant in Eq. (\ref{dP_dk_-_0}) represents a non-perturbative result, which cannot be obtained at any finite order of perturbation with respect to the field derivatives. \corr{The conditions in Eqs. (\ref{condition1})-(\ref{condition2}) define the domain of validity of the LCFA. It is interesting to notice that there exists a bound also on the second derivative of the field, which justifies our definition of $\tau(t)$ in Eq. (\ref{tau}). In addition, the conditions in Eqs. (\ref{condition1})-(\ref{condition2}) explicitly indicate that the LCFA becomes invalid for sufficiently small values of the ratio $k_-/p_-$, i.e., in the infrared region.}

The corresponding results for nonlinear Breit-Wheeler pair production are easily obtained by recalling the crossing-symmetry between this process and nonlinear Compton scattering \cite{Landau_b_4_1982,Baier_b_1998} (see also Ref. \cite{Di_Piazza_2017b}). In fact, the differential probability $dP_{BW}/dp_-$ of nonlinear Breit-Wheeler pair production per unit of produced positron light-cone energy $p_-$ is simply obtained from the corresponding probability $dP_C/dk_-$ of nonlinear Compton scattering by replacing $p_-\to -p_-$ ($\eta_0\to -(k_0p)/m^2$) and $k_-\to -k_-$, and then by multiplying the result by $-p_-^2/k_-^2$ \cite{Landau_b_4_1982,Baier_b_1998}. It is clear that in the case of nonlinear Breit-Wheeler pair production $k_-$ indicates the light-cone energy of the incoming photon. Thus, by writing $dP_{BW}/dp_-$ in the form $dP_{BW}/dp_-=\int d\varphi_+\,dP_{BW}/dp_-d\varphi_+$, we obtain
\begin{widetext}
\begin{gather}
\frac{dP_{BW}}{dp_-d\varphi_+}=\frac{\alpha}{2\pi}\frac{1}{k_-}\frac{1}{\theta_0}\text{Im}\int\frac{d\varphi_-}{\varphi_-+i0}\left[-1+\frac{p_-^2+p_-^{\prime\,2}}{4p_-p'_-}\bm{\Xi}^2_{\perp}(\varphi_-,\varphi_+)\right]e^{i\Phi_{BW}(p_-,\varphi_-,\varphi_+)},\\
\Phi_{BW}(p_-,\varphi_-,\varphi_+)=\frac{1}{2\theta_0}\frac{k^2_-}{p_-p'_-}\Bigg\{\varphi_-+\int_{-\varphi_-/2}^{\varphi_-/2}d\tilde{\varphi}\,\bm{\xi}^2_{\perp}(\varphi_++\tilde{\varphi})-\frac{1}{\varphi_-}\bigg[\int_{-\varphi_-/2}^{\varphi_-/2}d\tilde{\varphi}\,\bm{\xi}_{\perp}(\varphi_++\tilde{\varphi})\bigg]^2\Bigg\},
\end{gather}
\end{widetext}
where $\theta_0=(k_0k)/m^2$ and where here and below the quantity $p'_-=k_--p_-$ indicates the light-cone energy of the produced electron (as in the case of nonlinear Compton scattering the quantity $dP_{BW}/dp_-d\varphi_+$ can be rigorously interpreted as a differential probability only within the LCFA).

The corresponding differential probability $dP_{BW}^{\text{NLO}}/dp_-d\varphi_+$ up to the next-to-leading order in the LCFA is given by
\begin{widetext}
\begin{equation}
\label{dP_dp_-_NLO}
\begin{split}
&\frac{dP_{BW}^{\text{NLO}}}{dp_-d\varphi_+}=\frac{dP_{BW,\text{LCFA}}}{dp_-d\varphi_+}\\
&\quad-\frac{\alpha}{\sqrt{3}\pi}\frac{\theta_0}{k_-}\left\langle\frac{1}{45}\frac{\bm{\kappa}^{\prime\,2}(\varphi_+)+3\bm{\kappa}(\varphi_+)\cdot\bm{\kappa}^{\prime\prime}(\varphi_+)}{\kappa^4(\varphi_+)}\left[\frac{k^2_-}{p_-p'_-}\frac{1}{\kappa(\varphi_+)}\text{K}_{1/3}\left(\frac{2}{3}\frac{k^2_-}{p_-p'_-}\frac{1}{\kappa(\varphi_+)}\right)-2\text{K}_{2/3}\left(\frac{2}{3}\frac{k^2_-}{p_-p'_-}\frac{1}{\kappa(\varphi_+)}\right)\right]\right.\\
&-\frac{1}{45}\frac{\bm{\kappa}^{\prime\,2}(\varphi_+)+3\bm{\kappa}(\varphi_+)\cdot\bm{\kappa}^{\prime\prime}(\varphi_+)}{\kappa^4(\varphi_+)}\frac{p_-^2+p_-^{\prime\,2}}{p_-p'_-}\\
&\times\left\{6\text{K}_{2/3}\left(\frac{2}{3}\frac{k^2_-}{p_-p'_-}\frac{1}{\kappa(\varphi_+)}\right)-\left[\frac{k^2_-}{p_-p'_-}\frac{1}{\kappa(\varphi_+)}+4\frac{p_-p'_-}{k^2_-}\kappa(\varphi_+)\right]\text{K}_{1/3}\left(\frac{2}{3}\frac{k^2_-}{p_-p'_-}\frac{1}{\kappa(\varphi_+)}\right)\right\}\\
&\left.+\frac{1}{3}\frac{p_-^2+p_-^{\prime\,2}}{p_-p'_-}\frac{\bm{\kappa}(\varphi_+)\cdot\bm{\kappa}^{\prime\prime}(\varphi_+)}{\kappa^4(\varphi_+)}\left[\text{K}_{2/3}\left(\frac{2}{3}\frac{k^2_-}{p_-p'_-}\frac{1}{\kappa(\varphi_+)}\right)-\frac{p_-p'_-}{k^2_-}\kappa(\varphi_+)\text{K}_{1/3}\left(\frac{2}{3}\frac{k^2_-}{p_-p'_-}\frac{1}{\kappa(\varphi_+)}\right)\right]\right\rangle.
\end{split}
\end{equation}
\end{widetext}
where $\bm{\kappa}(\varphi_+)=\theta_0\bm{\xi}'_{\perp}(\varphi)$ ($\kappa(\varphi)=|\bm{\kappa}(\varphi)|$), and where
\begin{equation}
\begin{split}
\frac{dP_{BW,\text{LCFA}}}{dp_-d\varphi_+}=&\frac{\alpha}{\sqrt{3}\pi}\frac{1}{k_-}\frac{1}{\theta_0}\left[\frac{p_-^2+p_-^{\prime\,2}}{p_-p'_-}\right.\\
&\qquad\times\text{K}_{2/3}\left(\frac{2}{3}\frac{k^2_-}{p_-p'_-}\frac{1}{\kappa(\varphi_+)}\right)\\
&\left.+\int_{2k^2_-/3p_-p'_-\kappa(\varphi_+)}^{\infty} dz\,\text{K}_{1/3}(z)\right].
\end{split}
\end{equation}
It is important to notice that, since $k_-$ here is the fixed light-cone energy of the incoming photon, no ``infrared'' problems arise in the case of nonlinear Breit-Wheeler pair production. In fact, the quantity $k_-^2/p_-p'_-$ is always larger than or equal to four and cannot compensate large values of the parameter $\xi_0^2$ in the phase $\Phi_{BW}(p_-,\varphi_-,\varphi_+)$. Consequently, it does not make physical sense to compute the asymptotic expression as done for nonlinear Compton scattering in Eq. (\ref{dP_dk_-_NLO_s_km}). In addition, we would conclude that the systematic approach presented in this section is particularly useful in the case of nonlinear Breit-Wheeler pair production because the size of the correction does not depend significantly on the final energies of the particles and, provided that it is much smaller than the LCFA result, can be implemented without additional restrictions. By contrast, in the case of nonlinear Compton scattering the requirement that the correction has to be much smaller than the LCFA result restricts its applicability to the intermediate- and high-energy part of the emission spectrum.

\section{The LCFA for angular-resolved emission and pair-production probabilities}

As we have mentioned in the Introduction, the collinear approximation in nonlinear Compton scattering, i.e., the fact that the electron is assumed to emit along its instantaneous velocity, has been recently investigated in Ref.~\cite{Blackburn_2018}. In order to provide some analytical insight into this aspect of the LCFA, we start from Eq. (8) in Ref.~\cite{Di_Piazza_2018c}, which we can rewrite as
\begin{widetext}
\begin{equation}
\label{dP_d^3k}
\frac{dP_C}{d\bm{k}}=-\frac{\alpha}{4\pi^2}\frac{1}{\eta_0}\frac{1}{\omega_0\omega p'_-}\int d\varphi_+ d\varphi_-\,e^{i\frac{1}{2\eta_0}\frac{k_-}{p'_-}\int_{-\varphi_-/2}^{\varphi_-/2}d\tilde{\varphi}\,\left\{1+\left[\frac{\bm{p}_{\perp}}{m}-\frac{p_-}{k_-}\frac{\bm{k}_{\perp}}{m}-\bm{\xi}_{\perp}(\varphi_++\tilde{\varphi})\right]^2\right\}}\left[1+\frac{1}{4}\frac{p_-^2+p_-^{\prime\,2}}{p_-p'_-}\bm{\Xi}^2_{\perp}(\varphi_-,\varphi_+)\right].
\end{equation}
\end{widetext}

In the case of Eq. (\ref{dP_d^3k}), the LCFA amounts to expanding the function $\bm{\xi}_{\perp}(\varphi)$ for $\varphi$ around $\varphi_+$ up to linear terms in $\varphi_-$ in the pre-exponential function and up to quadratic terms in the phase, where finally only terms scaling as $\varphi_-^3$ are retained. By writing in general $dP/d\bm{k}$ as $dP/d\bm{k}=\int d\varphi_+\,dP/d\bm{k}d\varphi_+$, we obtain that within the LCFA
\begin{widetext}
\begin{equation}
\label{dP_d^3k_dphi_LCFA_i}
\frac{dP_{C,\text{LCFA}}}{d\bm{k}d\varphi_+}=-\frac{\alpha}{4\pi^2}\frac{1}{\eta_0}\frac{1}{\omega_0\omega p'_-}\\
\int d\varphi_-\,e^{i\frac{1}{2\eta_0}\frac{k_-}{p'_-}\left\{\left[1+\bm{\pi}_{\perp,-}^2(\varphi_+)\right]\varphi_-+\frac{\bm{\xi}^{\prime\,2}_{\perp}(\varphi_+)}{12}\varphi_-^3\right\}}\left[1+\frac{p_-^2+p_-^{\prime\,2}}{p_-p'_-}\frac{\bm{\xi}^{\prime\,2}_{\perp}(\varphi_+)}{4}\varphi_-^2\right],
\end{equation}
\end{widetext}
where in general
\begin{equation}
\bm{\pi}_{\perp,\pm}(\varphi_+)=\frac{\bm{p}_{\perp}}{m}-\frac{p_-}{k_-}\frac{\bm{k}_{\perp}}{m}\pm\bm{\xi}_{\perp}(\varphi_+).
\end{equation}
The remaining integral is easily carried out with the help of the formulas in the appendix and the final result is
\begin{widetext}
\begin{equation}
\label{dP_d^3k_dphi_LCFA}
\frac{dP_{C,\text{LCFA}}}{d\bm{k}d\varphi_+}=\frac{\alpha}{\sqrt{3}\pi^2}\frac{1}{\omega_0\omega p'_-}\frac{\sqrt{1+\bm{\pi}^2_{\perp,-}(\varphi_+)}}{\chi(\varphi_+)}\left\{\frac{p_-^2+p_-^{\prime\,2}}{p_-p'_-}\left[1+\bm{\pi}^2_{\perp,-}(\varphi_+)\right]-1\right\}\text{K}_{1/3}\left(\frac{2}{3}\frac{k_-}{p'_-}\frac{\left[1+\bm{\pi}^2_{\perp,-}(\varphi_+)\right]^{3/2}}{\chi(\varphi_+)}\right).
\end{equation}
\end{widetext}
Note that this result is in agreement with Eq. (4.13) in Ref.~\cite{Baier_b_1998}. The formation phase $\varphi_f$ of the emission of a photon with momentum between $\bm{k}$ and $\bm{k}+d\bm{k}$ can be defined from Eq. (\ref{dP_d^3k_dphi_LCFA_i}) in an analogous way as we have done in Ref.~\cite{Di_Piazza_2018c} and the result is obtained from Eq. (\ref{phi_f}) with the substitutions $|\bm{\xi}'_{\perp}(\varphi_+)|\to|\bm{\xi}'_{\perp}(\varphi_+)|/\sqrt{1+\bm{\pi}^2_{\perp,-}(\varphi_+)}$ and $\eta_0\to \eta_0/[1+\bm{\pi}^2_{\perp,-}(\varphi_+)]$ [note that $\chi(\varphi_+)\to \chi(\varphi_+)/[1+\bm{\pi}^2_{\perp,-}(\varphi_+)]^{3/2}$]. This implies that all considerations on the formation phase presented in the previous section and in Ref.~\cite{Di_Piazza_2018c} can be repeated for the differential probability $dP/d\bm{k}d\varphi_+$. Moreover, Eq. (\ref{dP_d^3k_dphi_LCFA}) indicates that the main contribution to the integral in $d^2\bm{k}_{\perp}$ comes from the region $[1+\bm{\pi}^2_{\perp,-}(\varphi_+)]^{3/2}\lesssim \chi(\varphi_+)p'_-/k_-$, such that at $\chi_0\sim 1$ and at a give phase $\varphi_+$ the instantaneous emission cone of low-energy photons with $p_-/\xi_0^3\ll k_-\ll p_-$ can be a factor $(p_-/k_-)^{1/3}$ broader than that of hard photons with $k_-\sim p_-$, which is of the order of $m/p_-$. 

Finally, for the sake of completeness, we report the corresponding differential probabilities $dP_{BW}/d^3\bm{p}$ and $dP_{BW,\text{LCFA}}/d^3\bm{p}d\varphi_+$ of non-linear Breit-Wheeler pair production, which can be obtained from Eqs. (\ref{dP_d^3k}) and (\ref{dP_d^3k_dphi_LCFA}), respectively, by means of the substitutions $p_-\to -p_-$ ($\eta_0\to -(k_0p)/m^2$), $\bm{p}_{\perp}\to -\bm{p}_{\perp}$, $k_-\to -k_-$, $\omega\to -\omega$, and  $\bm{k}_{\perp}\to -\bm{k}_{\perp}$ and by then multiplying by $-\omega p_-/\varepsilon k_-$:
\begin{widetext}
\begin{align}
\frac{dP_{BW}}{d\bm{p}}=&\frac{\alpha}{4\pi^2}\frac{1}{\theta_0}\frac{1}{\omega_0\varepsilon p'_-}\int d\varphi_+ d\varphi_-\,e^{i\frac{1}{2\theta_0}\frac{k^2_-}{p_-p'_-}\int_{-\varphi_-/2}^{\varphi_-/2}d\tilde{\varphi}\,\left\{1+\left[\frac{\bm{p}_{\perp}}{m}-\frac{p_-}{k_-}\frac{\bm{k}_{\perp}}{m}+\bm{\xi}_{\perp}(\varphi_++\tilde{\varphi})\right]^2\right\}}\left[1+\frac{1}{4}\frac{p_-^2+p_-^{\prime\,2}}{p_-p'_-}\bm{\Xi}^2_{\perp}(\varphi_-,\varphi_+)\right],\\
\frac{dP_{BW,\text{LCFA}}}{d\bm{p}d\varphi_+}=&\frac{\alpha}{\sqrt{3}\pi^2}\frac{1}{\omega_0\varepsilon p'_-}\frac{\sqrt{1+\bm{\pi}^2_{\perp,+}(\varphi_+)}}{\kappa(\varphi_+)}\left\{1+\frac{p_-^2+p_-^{\prime\,2}}{p_-p'_-}\left[1+\bm{\pi}^2_{\perp,+}(\varphi_+)\right]\right\}\text{K}_{1/3}\left(\frac{2}{3}\frac{k^2_-}{p_-p'_-}\frac{\left[1+\bm{\pi}^2_{\perp,+}(\varphi_+)\right]^{3/2}}{\kappa(\varphi_+)}\right),
\end{align}
\end{widetext}
where the momenta have to be interpreted according to discussion in the previous section (for example, $p'_-=k_--p_-$ is the light-cone energy of the produced electron).

\section{Conclusions}

In this paper we have developed a new scheme to implement numerically nonlinear Compton scattering beyond the LCFA by generalizing and simplifying the findings in Ref.~\cite{Di_Piazza_2018c}. We have provided a model which guarantees that the emission probability has the correct constant behavior in the infrared region of the spectrum, where the LCFA features an integrable divergence. These findings have been generalized from the special case of a plane wave to that of a general background electromagnetic field and can be implemented also in PIC codes in the sense that only local values of the physical quantities are employed.

In addition, we have determined the leading-order correction to the LCFA differential photon emission probability in the case of a plane wave by using a systematic approach, where the relative variation of the plane wave within the formation length is much smaller than unity. The expression of the correction indicates that the applicability of the LCFA also constrains the second derivative of the plane-wave field. Moreover, we have obtained the corresponding corrections to the LCFA differential positron spectrum in nonlinear Breit-Wheeler pair production. We have observed that in this case no infrared issues arise such that the correction obtained via the systematic approach is applicable for all positron light-cone energies.

Finally, for the sake of completeness, we have reported the fully differential photon emission spectrum and positron production spectrum in a plane wave within the LCFA.

\begin{acknowledgments}
SM was supported by the German Research Foundation (Deutsche Forschungsgemeinschaft, DFG) -- ME 4944/1-1. ADP and MT equally contributed to this work, being responsible of the analytical and of the numerical parts, respectively.
\end{acknowledgments}

\appendix

\section{Useful integrals}

In order to obtain the final expressions within the LCFA and beyond the following integrals are useful (see, e.g., Ref. \cite{Baier_b_1998}):
\begin{align}
&\int_{-\infty}^{\infty}\frac{dx}{x+i0}e^{ib(x+x^3/3)}=-\frac{2i}{\sqrt{3}}\int_{2b/3}^{\infty}dz\,\text{K}_{1/3}(z),\\
&\int_{-\infty}^{\infty}dx\, e^{ib(x+x^3/3)}=\frac{2}{\sqrt{3}}\text{K}_{1/3}\Big(\frac{2}{3}b\Big),\\
&\int_{-\infty}^{\infty}dx\, xe^{ib(x+x^3/3)}=\frac{2i}{\sqrt{3}}\text{K}_{2/3}\Big(\frac{2}{3}b\Big),\\
&\int_{-\infty}^{\infty}dx\, x^2e^{ib(x+x^3/3)}=-\frac{2}{\sqrt{3}}\text{K}_{1/3}\Big(\frac{2}{3}b\Big),\\
&\int_{-\infty}^{\infty}dx\, x^3e^{ib(x+x^3/3)}\\
&\qquad =-\frac{2i}{\sqrt{3}}\Big[\text{K}_{2/3}\Big(\frac{2}{3}b\Big)-\frac{1}{b}\text{K}_{1/3}\Big(\frac{2}{3}b\Big)\Big],\\
&\int_{-\infty}^{\infty}dx\, x^4e^{ib(x+x^3/3)}\\
&\qquad =\frac{2}{\sqrt{3}}\Big[\text{K}_{1/3}\Big(\frac{2}{3}b\Big)-\frac{2}{b}\text{K}_{2/3}\Big(\frac{2}{3}b\Big)\Big],\\
&\int_{-\infty}^{\infty}dx\, x^5e^{ib(x+x^3/3)}\\
&\qquad =\frac{2i}{\sqrt{3}}\Big[\text{K}_{2/3}\Big(\frac{2}{3}b\Big)-\frac{6}{b}\text{K}_{1/3}\Big(\frac{2}{3}b\Big)\Big],\\
&\int_{-\infty}^{\infty}dx\, x^6e^{ib(x+x^3/3)}\\
&\qquad =\frac{2}{\sqrt{3}}\Big[\frac{6}{b}\text{K}_{2/3}\Big(\frac{2}{3}b\Big)-\text{K}_{1/3}\Big(\frac{2}{3}b\Big)\Big(1+\frac{4}{b^2}\Big)\Big],
\end{align}
where $b$ is a positive real number and $\text{K}_{\nu}$ indicates the modified Bessel function of order $\nu$ \cite{NIST_b_2010}.

\end{document}